\documentclass[conference]{IEEEtran}
\IEEEoverridecommandlockouts
\usepackage{cite}
\usepackage{amsmath,amssymb,amsfonts}
\usepackage{algorithmic}
\usepackage{graphicx}
\usepackage{textcomp}
\usepackage{xcolor}
\def\BibTeX{{\rm B\kern-.05em{\sc i\kern-.025em b}\kern-.08em
    T\kern-.1667em\lower.7ex\hbox{E}\kern-.125emX}}
\usepackage{amssymb}  
\usepackage{pifont}   
\newcommand{\xmark}{\ding{55}}  
\usepackage{array} 

\begin{document}

\title{AI-Protected Blockchain-based IoT environments: Harnessing the Future of Network Security and Privacy

}

\author{Ali Mohammadi Ruzbahani\\
University of Calgary\\
\textit{Email: ali.mohammadiruzbaha@ucalgary.ca}
}

\maketitle

\begin{abstract}
Integrating blockchain technology with the Internet of Things offers transformative possibilities for enhancing network security and privacy in the contemporary digital landscape, where interconnected devices and expansive networks are ubiquitous. This paper explores the pivotal role of artificial intelligence in bolstering blockchain-enabled IoT systems, potentially marking a significant leap forward in safeguarding data integrity and confidentiality across networks. Blockchain technology provides a decentralized and immutable ledger, ideal for the secure management of device identities and transactions in IoT networks. When coupled with AI, these systems gain the ability to not only automate and optimize security protocols but also adaptively respond to new and evolving cyber threats. This dual capability enhances the resilience of networks against cyber-attacks, a critical consideration as IoT devices increasingly permeate critical infrastructures. The synergy between AI and blockchain in IoT is profound. AI algorithms can analyze vast amounts of data from IoT devices to detect patterns and anomalies that may signify security breaches. Concurrently, blockchain can ensure that data records are tamper-proof, enhancing the reliability of AI-driven security measures. Moreover, this research evaluates the implications of AI-enhanced blockchain systems on privacy protection within IoT networks. IoT devices often collect sensitive personal data, making privacy a paramount concern. AI can facilitate the development of new protocols that ensure data privacy and user anonymity without compromising the functionality of IoT systems.
Through comprehensive analysis and case studies, this paper aims to provide an in-depth understanding of how AI-enhanced blockchain technology can revolutionize network security and privacy in IoT environments.
\end{abstract}


\section{Introduction}
The modern digital landscape is marked by the integration of innovative technologies that have fundamentally altered various aspects of human life and industry \cite{h6,k1}. The advent of blockchain technology has been particularly transformative, providing a decentralized platform that enhances security and trust without the need for a central authority\cite{1,h7,k1}. Its core characteristics—decentralization, transparency, security, and immutability—have been widely recognized for their potential to revolutionize multiple sectors \cite{h1,2}. Blockchain technology's evolution from powering cryptocurrencies to enabling complex decentralized applications (DApps) through distributed ledgers has been significant \cite{h2,4,h8,k2}. This technology is now poised to play a crucial role in the Internet of Things (IoT) by ensuring more secure and private connections between devices. The combination of blockchain with artificial intelligence (AI) in IoT environments presents an exciting frontier \cite{h9,5,k3}. AI's capability to enhance blockchain operations through intelligent, automated, and adaptive security responses can significantly increase the resilience of IoT networks against cyber threats \cite{3,k4,h10}.

The focus of this paper is to explore how AI can be integrated with blockchain technology to develop robust security solutions that ensure both the security and privacy of data in IoT frameworks \cite{6,h11}. This involves a detailed examination of AI-driven security protocols, innovative data privacy measures, and the overall implications of these technologies on the digital ecosystem. The integration of AI with blockchain in IoT not only promises enhanced security features but also opens up new avenues for privacy preservation in interconnected devices \cite{7,k5}. Through this study, we aim to provide insights and strategic guidance on implementing AI-protected blockchain-based systems in IoT, contributing to the advancement of network security and privacy in an increasingly digital world \cite{8,h12,9}. AI applications in this context serve multiple functions. They can automate complex processes, analyze large streams of IoT data to detect anomalies and potential threats, and optimize the performance and efficiency of blockchain networks. Machine learning algorithms, a subset of AI, are particularly adept at pattern recognition, which can be applied to predict and counteract emerging security threats in real-time \cite{k6}. Additionally, AI can manage and enforce blockchain's consensus protocols more dynamically, ensuring a more responsive and adaptive network \cite{10,h13}.

Privacy concerns in IoT are paramount as these devices often collect and transmit personal data across networks. Integrating blockchain and AI can enhance privacy through mechanisms such as smart contracts and advanced encryption techniques. These tools can automate the execution of agreements and protect data transactions, ensuring they are secure from unauthorized access and tampering \cite{k7}. Furthermore, AI can enhance blockchain's capability to offer decentralized privacy-preserving technologies such as zero-knowledge proofs, allowing IoT devices to validate transactions without revealing any underlying personal data \cite{11,h14}. The potential for these integrated technologies to transform IoT security and privacy extends into various industries, including healthcare, automotive, and smart homes, where the integrity and confidentiality of data are crucial. In healthcare, for example, ensuring the privacy and security of patient data while utilizing IoT devices for monitoring and treatment can significantly enhance patient trust and treatment outcomes \cite{12,h3}. Similarly, in the automotive sector, blockchain and AI can secure vehicular communication systems, making connected vehicles safer from cyber-attacks and privacy breaches \cite{13}.

In conclusion, the integration of AI and blockchain technology holds the potential to significantly fortify IoT environments against security threats and privacy issues. This paper has explored the synergistic effects of these technologies and proposed a framework for their implementation in IoT systems. Moving forward, it is essential for industry stakeholders and regulators to consider these findings in their strategic planning and policy-making to ensure the secure and ethical use of IoT technologies in a digitally connected world \cite{14,h4,15,h15}.
\begin{table*}[h]
\centering
\caption{Comparison of Existing Surveys on AI-Enhanced Blockchain in IoT}
\begin{tabular}{|c|c|c|c|c|>{\centering\arraybackslash}p{4cm}|}
\hline
Reference                   & Investigated Challenges & Role of Blockchain & Role of AI & Security Enhancement & Remark\\
\hline
\cite{Ex-Surv-Jour001}      & \checkmark             & \checkmark         & \xmark     & \xmark                & Focused on security challenges in blockchain applications.\\
\cite{Ex-Surv-Jour003}      & \checkmark             & \checkmark         & \xmark     & \xmark                & Reviewed AI applications for threat detection in IoT.\\
\cite{Ex-Surv-Jour004}      & \xmark                 & \checkmark         & \checkmark & \xmark                & Discussed AI-driven blockchain architectures for IoT.\\
\cite{Ex-Surv-Jour005}      & \xmark                 & \checkmark         & \checkmark & \checkmark            & Explored AI-enhanced security solutions in blockchain IoT frameworks.\\
\cite{Ex-Surv-Jour006}      & \checkmark             & \xmark             & \checkmark & \xmark                & Analyzed challenges in integrating AI with blockchain for IoT.\\
\hline
\end{tabular}
\label{ta1}
\end{table*}

\subsection{Existing Surveys}
As the integration of blockchain technology and artificial intelligence (AI) continues to redefine the landscape of Internet of Things (IoT) environments, scholarly interest has increasingly focused on how these technologies can collectively enhance network security and privacy \cite{k4,h5,k6}. Existing research has typically examined the capabilities and challenges of blockchain and AI within IoT systems in isolation. For example, numerous studies have investigated how the inherent security features of blockchain can safeguard IoT devices and data, while others have scrutinized AI's role in real-time detection and response to security threats \cite{10,h16,11,t6}.
A pivotal systematic literature review highlighted the current applications and security solutions provided by blockchain in IoT, emphasizing the need for advanced mechanisms to counter dynamic and sophisticated cyber threats \cite{Ex-Surv-Jour001}. Concurrently, another significant study assessed advancements in AI algorithms capable of autonomously detecting anomalies and potential breaches in IoT networks, proposing frameworks for integrating these algorithms with blockchain technologies \cite{Ex-Surv-Jour003}.
Further research has proposed architectural frameworks that meld AI and blockchain to boost the security and efficiency of IoT systems \cite{Ex-Surv-Jour004}. These frameworks strive to capitalize on AI's predictive capabilities and blockchain's tamper-resistant properties to forge a secure and resilient digital infrastructure.

Moreover, surveys such as \cite{Ex-Surv-Jour005} have delved into multi-layered security approaches that amalgamate the strengths of AI and blockchain. These approaches present comprehensive protection strategies for interconnected devices across various sectors, outlining the layers of security that can be enhanced through the integration of AI and blockchain.
However, in contrast to the existing literature that often treats AI and blockchain as distinct and separate entities, our study takes a novel approach by exploring their synergistic integration within IoT environments. This paper aims to offer a holistic view of how combined AI and blockchain technologies can revolutionize network security and privacy, marking a significant advancement in the field \cite{13,12,h17}.
By leveraging AI's machine learning capabilities to analyze data patterns and predict potential threats, and combining this with blockchain's ability to secure data through its immutable ledger system, we can create a robust framework for IoT security. This integration not only enhances the detection of anomalies but also ensures the integrity and confidentiality of the data being transmitted across IoT devices.

Furthermore, the paper discusses the implications of this integrated approach on privacy concerns, a critical aspect as IoT devices pervasively collect and process personal data. The use of blockchain can provide a decentralized approach to data management, reducing the risks associated with central data storage, while AI can enhance data processing and control access, ensuring that data privacy is maintained without sacrificing efficiency.
In addition to technical enhancements, our study also addresses the policy and regulatory challenges that arise with the deployment of AI and blockchain in IoT. It advocates for the development of clear guidelines and standards to govern the ethical use of these technologies, ensuring that they contribute positively to societal needs while safeguarding against misuse.

In conclusion, the synergy between AI and blockchain holds the potential to transform IoT security and privacy fundamentally. Our research not only maps out the technological advancements but also provides strategic guidance for future implementations, aiming to foster an environment where technology serves as a cornerstone for secure, efficient, and trustworthy digital interactions.

\subsection{Contributions}
This survey makes significant contributions to the domain of network and service security by focusing on the integration of AI with blockchain technology within IoT environments. We address a gap in the current literature by providing an exhaustive analysis of how AI can enhance blockchain applications in IoT, ensuring more secure and private networks. Our study distinguishes itself by examining a broad spectrum of security mechanisms and vulnerabilities specific to IoT environments, exploring how AI-enhanced blockchain solutions can address these challenges.

By evaluating real-world applications and potential future developments, this survey serves as a vital resource for both researchers and practitioners. It offers a comprehensive understanding of the current state of IoT security, the role of AI in enhancing these frameworks, and the transformative potential of blockchain technology in creating more robust digital networks.

\subsection{Paper Organization}
The structure of this survey begins with Section \ref{s2}, which outlines the methodological approach and presents the key research questions. Section \ref{s21} investigates the current landscape of IoT security challenges and how blockchain technology can address them. Section \ref{s3} delves into AI's role in enhancing security protocols within blockchain-based IoT systems. Section \ref{s4} explores the practical applications and security frameworks that combine AI and blockchain in IoT. Section \ref{s5} discusses the implications of these technologies on future network security and privacy. The final section, \ref{s6}, synthesizes the findings and discusses potential directions for future research in AI-protected blockchain-based IoT environments. Concluding remarks are provided in Section \ref{ff8}.

\begin{table}[]
\centering
\caption{Cyber Attacks on AI-Enhanced Blockchain in IoT}
\footnotesize
\begin{tabular}{|>{\centering\arraybackslash}p{1.8cm}|>{\centering\arraybackslash}p{1.8cm}|>{\centering\arraybackslash}p{1.8cm}|>{\centering\arraybackslash}p{1.8cm}|}
\hline
\textbf{Attack Type} & \textbf{Method} & \textbf{Impact} & \textbf{Objective} \\
\hline
Network Intrusions & Unauthorized entry & Undermines network integrity & Seeks unauthorized access and control \\
\hline
Data Breaches & Leveraging security flaws & Reveals confidential data & Aims to steal or alter private information \\
\hline
Man-in-the-Middle Attacks & Interfering with data flows & Modifies or intercepts data during transmission & Targets the acquisition and alteration of sensitive data \\
\hline
Ransomware Attacks & Holding data hostage for ransom & Disrupts operations & Demands payment in exchange for access \\
\hline
AI Poisoning & Corrupting machine learning data & Misdirects AI decision-making & Seeks to alter AI operations to benefit the attacker \\
\hline
Side Channel Attacks & Utilizing indirect system information & Accesses critical operational data & Obtains sensitive data through indirect means \\
\hline
Sybil Attacks & Creating numerous fictitious identities & Compromises consensus mechanisms & Affects decision-making or alters transaction records \\
\hline
Distributed Denial of Service (DDoS) Attacks & Bombarding the network with excessive requests & Overwhelms system availability & Halts IoT functions and services \\
\hline
\end{tabular}
\label{ta2}
\end{table}

\section{Research Methodology and Questions}\label{s2}
Our research methodology rigorously investigates the integration of artificial intelligence (AI) with blockchain technology in IoT settings, focusing on boosting network security and privacy. We adopt a qualitative research strategy, compiling and analyzing data from a broad range of academic papers and formulating specific metrics to address several critical research questions. This approach enables us to assess the current landscape of blockchain and AI utilization in IoT, explore the integration challenges they face, and identify potential strategies for enhancing these technologies to create stronger security frameworks \cite{k7,h18,13}. The research is guided by the following key questions:

\begin{itemize}
\item \textbf{RQ1:} What are the existing uses of blockchain and AI in strengthening IoT security?
\item \textbf{RQ2:} What are the main challenges and solutions associated with merging AI with blockchain technology in IoT contexts?
\item \textbf{RQ3:} In what ways can AI amplify the effectiveness of blockchain-driven security mechanisms in IoT?
\end{itemize}

To address these questions, we have identified pertinent research papers, including journal articles, conference proceedings, and other publications that describe different techniques, frameworks, and methodologies used in enhancing IoT security through blockchain and AI integration. The subsequent sections discuss the findings for each research question.

\section{Cyber Attacks on IoT Systems}\label{s21}
IoT systems, while transformative, are susceptible to a range of cyber threats that underscore the need for advanced security measures. These threats include network intrusions, data breaches, and other sophisticated cyber attacks that exploit vulnerabilities in IoT devices and networks\cite{k8}. AI-enhanced blockchain solutions offer promising avenues for defending against these threats by providing decentralized security mechanisms and intelligent threat detection and response systems. This section outlines various types of cyber attacks common in IoT environments and discusses how blockchain and AI technologies can mitigate these risks effectively\cite{14,15}.

\section{Blockchain and AI Integration in IoT Security}\label{s3}
This section explores the dynamic field of blockchain and AI integration within IoT security frameworks. We highlight studies that apply both technologies to create innovative solutions for securing IoT devices and data. This includes AI-driven anomaly detection systems that utilize blockchain for data integrity and decentralized operations, enhancing the trustworthiness and reliability of IoT networks. The exploration of blockchain and AI in this context illuminates the vast potential of these technologies to provide comprehensive security solutions that are both scalable and robust.

\section{Enhancing IoT Security with AI-Protected Blockchain Solutions}\label{s4}
The integration of AI and blockchain holds significant promise for enhancing IoT security. This section discusses various applications of these technologies in IoT, such as secure communications, data privacy, and device authentication. By leveraging AI's ability to quickly analyze patterns and detect anomalies, combined with blockchain's capabilities for ensuring data integrity and preventing tampering, these integrated solutions address key security challenges in IoT. We review practical implementations and theoretical models that illustrate how AI-protected blockchain solutions can effectively secure IoT environments, fostering greater adoption and trust in these technologies\cite{k9}.

\begin{figure}[!ht]
	\centerline{\includegraphics[width=20pc]{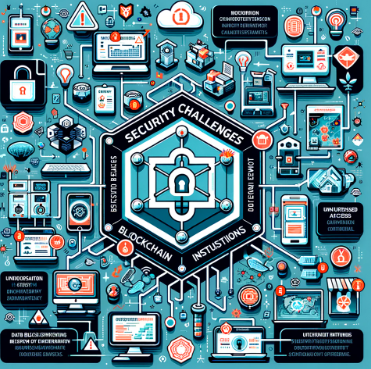}}
	\caption{Overview of security challenges of SC}
	\label{f1}
\end{figure}

\section{Role of AI-Enhanced Blockchain in IoT Network and Service Security}\label{s5}
This subsection explores various instances where blockchain, enhanced by artificial intelligence (AI), has been leveraged to bolster network and service security, particularly within IoT environments. An example is the AI-Blockchain Hybrid System, which integrates AI's predictive and adaptive capabilities with blockchain's security features to enhance network resilience and data integrity\cite{14,15}. This approach is designed to efficiently handle dynamic security threats and optimize IoT device performance through real-time data analysis and decentralized management.

In the context of improving security protocols, the study \cite{Sec-SC-Conf034} introduces an AI-driven monitoring system that employs blockchain for secure data logging and anomaly detection in network transactions. This system provides a robust framework for preventing and responding to cyber threats in a decentralized manner, reducing the risk of single points of failure\cite{h18, 16}.

Another crucial aspect is the use of AI to enhance blockchain's operational efficiency and security in diverse sectors. For instance, in healthcare, \cite{khan2022biomt,17} discusses a blockchain-based system, powered by AI, to secure health-related transactions and manage patient data across decentralized networks. This ensures enhanced privacy, security, and interoperability within healthcare systems.

Furthermore, in cloud environments, \cite{tan2022novel,18} presents a novel AI-Blockchain architecture for cloud manufacturing that uses intelligent algorithms to optimize Service Level Agreements (SLAs) and ensure trust among different cloud service stakeholders. This model not only enhances security but also improves the scalability and efficiency of cloud services.

These examples showcase the diverse applications of AI-enhanced blockchain technology in strengthening network and service security across various domains, particularly in IoT settings\cite{k9,k10,19}.

\section{Future Roadmap: The Convergence of AI and Blockchain in Enhancing Digital Security}\label{s6}
\begin{figure}[!ht]
	\centerline{\includegraphics[width=20pc]{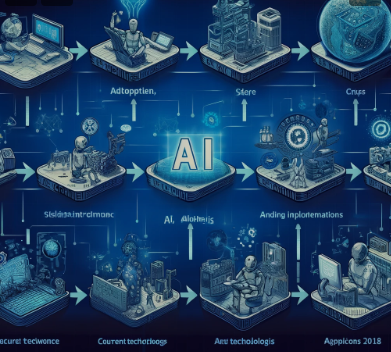}}
	\caption{The evolution of adopting AI in SC}
	\label{f2}
\end{figure}

As depicted in Fig. \ref{f2}, the integration of AI with blockchain has progressively evolved, enhancing the capabilities and efficiency of blockchain applications in security-sensitive environments such as IoT. This roadmap illustrates the transformative impact of AI in enabling blockchain systems to adapt and respond to emerging security challenges dynamically.

The potential for AI to optimize blockchain technology for security purposes is vast. Researchers are developing sophisticated AI tools to improve anomaly detection and security management within blockchain networks. For instance, the machine-learning-based tool, MLBP-Blockchain, discussed in \cite{Prom-of-AI-Conf008}, predicts and identifies potential security flaws in blockchain applications, enabling proactive security measures.

Further exploration into AI-driven security solutions for blockchain includes techniques such as those presented in \cite{Prom-of-AI-Earl-Jour001} and \cite{Prom-of-AI-Conf003,20}, which utilize deep learning and graph neural networks, respectively, to enhance the detection of vulnerabilities in blockchain networks. These advancements provide more precise and efficient security assessments.

Looking forward, the intersection of AI and blockchain is expected to yield innovative solutions that not only enhance the security and efficiency of digital networks but also support new applications, such as intelligent automated systems in smart cities, secure supply chain management, and privacy-preserving data sharing\cite{22,h18}. Such developments promise to revolutionize network and service security, paving the way for a more secure and interconnected digital future\cite{k10,k11}.

\section{The Influence of AI on Blockchain Technology for IoT Security}\label{s5}
\begin{figure}[!ht]
\centerline{\includegraphics[width=18pc]{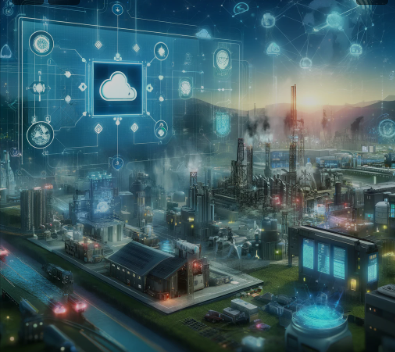}}
\caption{Future trends of AI in Blockchain for IoT Security}
\label{f3}
\end{figure}

This section delves into the ongoing research into the influence of AI on blockchain technology, specifically within IoT environments, as illustrated in Fig. \ref{f3}. Key trends include AI-enhanced security protocols, autonomous blockchain management, AI-driven dispute resolution, and the incorporation of bio-inspired AI algorithms. These topics are further explored in subsections \ref{sub1}, \ref{sub2}, \ref{sub3}, and \ref{sub4}, respectively.

\subsection{AI-Enhanced Security Protocols}\label{sub1}
Artificial Intelligence (AI) is revolutionizing security management within Internet of Things (IoT) networks by integrating enhanced capabilities through machine learning and natural language processing. These AI tools allow for more efficient and accurate security monitoring, significantly reducing human error and accelerating response times. In blockchain-based IoT security systems, the potential of AI is particularly vast. AI technologies can automate and optimize threat detection and response processes by continuously learning from network activities and emerging threats. This shift toward AI-enabled solutions initiates a new era of intelligent security management where systems can predict and mitigate risks autonomously. Furthermore, AI can tailor security measures based on the behavior patterns and vulnerability assessments of the network, ensuring a proactive stance against cyber threats. This not only enhances the security of IoT devices but also ensures the integrity and confidentiality of the data flowing through these networks. By leveraging AI, blockchain technology can be transformed into a more dynamic and secure framework that supports the growing complexity and scale of IoT ecosystems.

\subsection{Autonomous Blockchain Management}\label{sub2}
Drawing parallels from the finance sector, where AI-driven robo-advisors have revolutionized investment management, similar AI algorithms could be adapted to autonomously manage blockchain operations. This adaptation would enhance the efficiency and security of blockchain applications in IoT settings. Autonomous AI systems would actively monitor blockchain performance, dynamically adjust protocols, and respond to threats in real-time without human intervention. This level of automation democratizes advanced security strategies, previously only accessible to large-scale operators, and extends them to smaller entities or individual IoT applications. The implementation of such autonomous systems could lead to significant improvements in the resilience and reliability of blockchain networks, ensuring seamless and secure operations across various IoT applications. This shift not only streamlines the management of blockchain networks but also opens up new possibilities for the broader adoption and scalability of blockchain technology in diverse sectors.

\subsection{AI-Driven Dispute Resolution}\label{sub3}
AI's application in Online Dispute Resolution (ODR) systems exemplifies its potential to facilitate quicker and fairer resolutions in blockchain transactions, especially relevant in IoT environments. By analyzing transaction data and contractual conditions, AI can provide objective analyses and predict likely outcomes, thereby expediting the dispute resolution process within decentralized networks. This application of AI in dispute resolution not only enhances the efficiency of resolving conflicts but also increases the transparency and trust in blockchain transactions. Automated, AI-driven solutions can handle a high volume of disputes simultaneously, reducing the need for human arbitrators and decreasing the overall time and cost associated with dispute resolutions. Such capabilities are crucial in IoT scenarios where transactions occur frequently and disputes need to be resolved swiftly to maintain the integrity of the network.

\subsection{Bio-Inspired Blockchain Solutions}\label{sub4}
Bio-inspired AI technologies, such as neural networks and genetic algorithms, present exciting opportunities for enhancing blockchain applications, particularly in managing complex, dynamic tasks in IoT security. For instance, swarm intelligence could be utilized to optimize resource allocation across blockchain networks, improving efficiency and scalability. Similarly, genetic algorithms can be applied to evolve blockchain protocols over time, enhancing their adaptability and robustness. The incorporation of bio-inspired AI models into blockchain frameworks allows for the simulation of natural processes that can significantly improve decision-making and problem-solving capabilities within these systems. Such advancements not only bolster the security features of blockchain but also enhance its capacity to adapt to changing network conditions and emerging threats in IoT environments.

\section{Conclusion}\label{ff8}
In conclusion, our research underscores the transformative role of AI in enhancing blockchain technology for the security of the IoT. AI not only strengthens existing security protocols but also introduces innovative capabilities such as autonomous management and adaptive dispute resolution mechanisms. As the integration of AI with blockchain deepens, the potential for these technologies to comprehensively safeguard and optimize IoT ecosystems becomes increasingly significant.
This paper has thoroughly explored how AI can dynamically adapt to and effectively address the continuously evolving security challenges in decentralized networks. By leveraging AI's advanced analytical and predictive capabilities, blockchain systems can become not only more secure but also smarter, thereby promising a future where digital transactions are both secure and intelligent. The integration of AI with blockchain offers a dual advantage: AI's ability to swiftly analyze and respond to threats complements blockchain's inherent strengths in providing a tamper-proof and transparent transaction ledger.
Moreover, the findings of this study offer a solid foundation for future research and development in this area. They suggest a clear path toward the creation of an integrated, adaptive, and robust digital infrastructure for IoT security that can effectively counteract and mitigate potential cyber threats. This investigation highlights the importance of such advancements in the context of increasing reliance on IoT technologies across various sectors, including healthcare, finance, and smart cities.
Furthermore, this research serves as a roadmap for both researchers and practitioners in the field. It encourages further exploration into the potential of AI-enhanced blockchain technologies, aiming to leverage these advancements to secure digital networks and systems comprehensively. By continuing to explore these technologies, we can unlock powerful tools for enhancing data integrity, confidentiality, and operational efficiency in an increasingly interconnected world.
Ultimately, the integration of AI with blockchain represents a significant step forward in our approach to IoT security. It calls for ongoing innovation and collaboration among technologists, industry leaders, and policymakers to ensure that the deployment of these technologies maximizes benefits while minimizing risks to privacy and security.

\bibliographystyle{IEEEtran}
\bibliography{References}
\end{document}